\documentclass{iagproc}

\usepackage{graphics}
\usepackage{graphicx}

\title{GINS: a new tool for VLBI Geodesy and Astrometry}
\author{G. Bourda$^{1}$, P. Charlot$^{1}$, R. Biancale$^{2}$                                          \\ 
        (1) Observatoire Aquitain des Sciences de l'Univers - Universit\'{e} Bordeaux I               \\
	    Laboratoire d'Astrophysique de Bordeaux - UMR5804/CNRS - Floirac, France                  \\
	(2) Centre National d'Etudes Spatiales - Groupe de Recherche de G\'{e}od\'{e}sie Spatiale     \\
	    Toulouse, France                                                                          \\}

\begin{document}

\maketitle

\begin{abstract}
In the framework of the "Groupe de Recherches de G\'{e}od\'{e}sie Spatiale" (GRGS), a rigorous combination of the data from five space geodetic techniques (VLBI, GPS, SLR, LLR and DORIS) is routinely applied to simultaneously determine a Terrestrial Reference Frame (TRF) and Earth Orientation Parameters (EOP). This analysis is conducted with the software package GINS which has the capability to process data from all five techniques together. Such a combination at the observation level should ultimately facilitate fine geophysical studies of the global Earth system. In this project, Bordeaux Observatory is in charge of the VLBI data analysis, while satellite geodetic data are processed by other groups. In this paper, we present (i) details about the VLBI analysis undertaken with GINS, and (ii) the results obtained for the EOP during the period 2005--2006. We also compare this EOP solution with the IVS (International VLBI Service for geodesy and astrometry) analysis coordinator combined results. The agreement is at the 0.2 mas level, comparable to that of the other IVS analysis centers, which demonstrates the capability of the GINS software for VLBI analysis.
\end{abstract}

\begin{keywords}
astrometry, geodesy, reference systems, Earth rotation, VLBI 
\end{keywords}

%--------------------------------------------------------------------------------------
\section{Introduction}

The software package GINS (G\'{e}od\'{e}sie par Int\'{e}grations Nu\-m\'{e}riques Simultan\'{e}es) is a multi-technique software initially developed by the GRGS/CNES (Groupe de Recherches de G\'{e}od\'{e}sie Spatiale" -- Centre National d'Etudes Spatiales, Toulouse, France) for analysing satellite geodetic data, and extended at later stages for analysing data from other space geodetic techniques \citep{Meyer2000}. Currently, GPS (Global Positionning System), DORIS (Doppler Orbitography and Radio-positioning Integrated by Satellite), SLR (Satellite Laser Ranging), LLR (Lunar Laser Ranging) and VLBI (Very Long Baseline Interferometry) observations can be processed with GINS. The parameters that can be estimated comprise satellite orbits around the Earth or another body of the solar system, gravity field coefficients, Earth Orientation Parameters (EOP), station coordinates, or other geophysical parameters. The well-known GRIM5 and EIGEN gravity field models were produced with GINS in particular \citep{Biancale2000,Reigber2002}. 
 
\begin{figure*}[htbp]
\centering
  \includegraphics[scale=0.65]{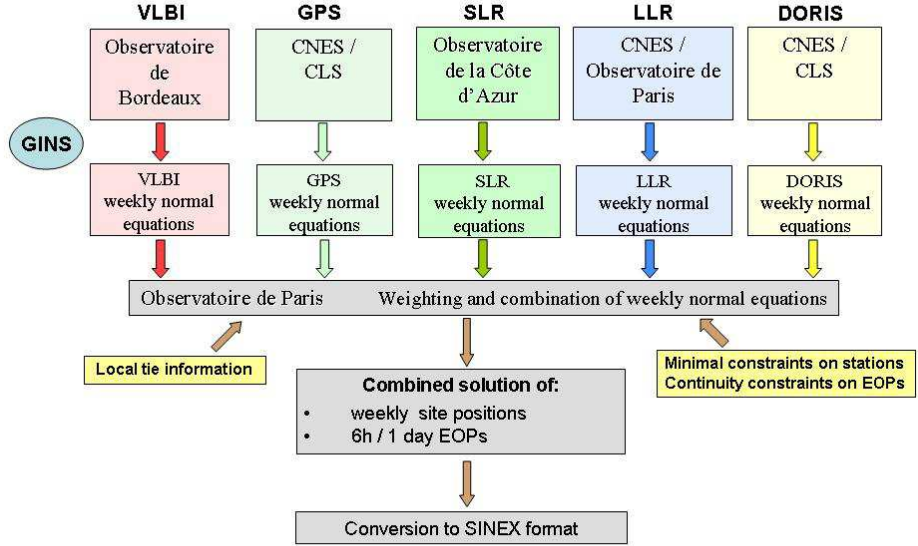}
\caption{Organization of the coordinated project of the GRGS for multi-technique combination at the observation level. CLS (\textit{Collecte Localisation Satellite}) is a private company funded in particular by the CNES.}
\label{fig:CRC_Project}
\end{figure*}

A rigorous combination of all the above space geodetic data has been developed to estimate station coordinates and EOP simultaneously from all techniques in the framework of the IERS (International Earth Rotation and Reference Systems Service) multi-technique combination pilot project. 
In this analysis, observations of the different astro-geodetic techniques (VLBI, GPS, SLR, LLR and DORIS) are first processed separately using GINS. The weekly datum-free normal equation matrices derived from the analyses of the different techniques are then combined to estimate station coordinates and EOP \citep{Coulot2007,Gambis2007}. Results are made available to the IERS in the form of SINEX files. 
In this project, the VLBI data are analysed in Bordeaux, while the satellite geodetic data are processed either in Toulouse (for GPS, DORIS and LLR) or Grasse (for SLR), with the final combination produced at Paris Observatory (see Figure~\ref{fig:CRC_Project} for more details about the project organization). 

The strength of the method is in the use of a unique software for all techniques with identical and up-to-date models and standards, ensuring homogeneous and reliable combined products. In addition, the solution benefits from complementary constraints brought by the various techniques. 

In this paper, we present an overview of the analyses undertaken with this new VLBI software, the results obtained for the EOP from 2005 to 2006, and the comparisons made with the IVS (International VLBI Service for geodesy and astrometry) analysis coordinator combined results. These comparisons indicate that GINS is at the level of the other VLBI analysis software packages.

%--------------------------------------------------------------------------------------
\section{VLBI analysis with GINS: data and modeling}

Since 2005 the regular weekly VLBI data acquired by the IVS have been routinely processed with the GINS software in order to estimate the EOP and the VLBI station positions. These data include both the IVS intensive sessions (i.e. one-hour long daily experiments) and the so-called IVS-R1 and IVS-R4 sessions (i.e. two 24-hour experiments per week). Overall, a total of 20~stations have been used in such sessions. 

Based on these data, weekly normal matrices are produced for combination with the data acquired by the other space geodetic techniques (GPS, SLR, LLR and DORIS). The free VLBI parameters include station positions and the five EOP ($X_p$, $Y_p$, $UT1-UTC$, d$\psi$, d$\epsilon$) along with clock and troposphere parameters. The clocks are modeled using piecewise continuous linear functions with breaks every two hours. The tropospheric zenith delays are modeled in a similar way except that breaks are applied every hour. Continuity constraints of 10~$\mu$s and 10~cm are applied to the clock and troposphere breaks, respectively. The a priori EOP series used is the IERS C04 series, the a priori Terrestrial Reference Frame (TRF) is VTRF2005 \citep{Nothnagel2005}, while the celestial frame is fixed to the ICRF (International Celestial Reference Frame; \citealp{Ma1998,Fey2004}). 

The following tidal and atmospheric models are also used in the analysis:
\begin{itemize}
\item  IERS~Conventions~2003 for solid Earth tides and pole tide models \citep{McCarthy2003},
\item  FES2004 for oceanic tides and oceanic loading models \citep{Lyard2006},
\item  6h-ECMWF (European Center for Meteorological Weather Forecast) atmospheric pressure fields only over continents (inverse barometer hypothesis) for atmospheric loading model, 
\item  Niell tropospheric mapping functions \citep{Niell1996}.
\end{itemize}

%--------------------------------------------------------------------------------------
\section{VLBI analysis with GINS: results and comparison}

In this section, we present the VLBI-only EOP results obtained based on a fixed TRF (VTRF2005) and compare these to the IVS combined EOP series. One set of EOP ($X_p$, $Y_p$, $UT1-UTC$, d$\psi$, d$\epsilon$) was estimated for every 24-hour session.

\begin{figure}
\centering
  \includegraphics[scale=0.3,angle=-90]{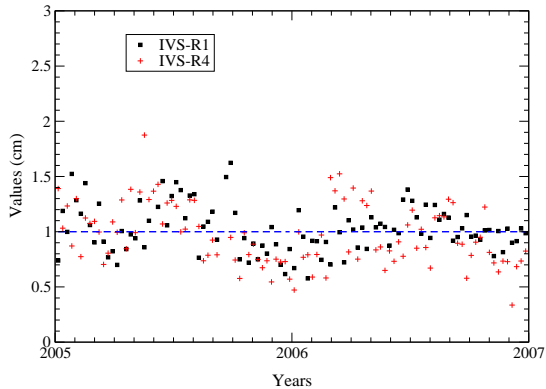}
\caption{Post-fit weighted RMS delay residuals with GINS for the  IVS-R1 and IVS-R4 sessions conducted in 2005-2006.}
\label{fig:Residuals}
\end{figure}

Figure~\ref{fig:Residuals} shows the post-fit weighted RMS (Root Mean Square) delay residuals obtained with GINS for the IVS-R1 and IVS-R4 sessions in 2005--2006. The RMS average over this period is 1.08 cm (i.e. 36 ps) for the IVS-R1 sessions, and 0.97 cm (i.e. 32 ps) for the IVS-R4 sessions.

\begin{figure}
\centering
  \includegraphics[scale=0.3,angle=-90]{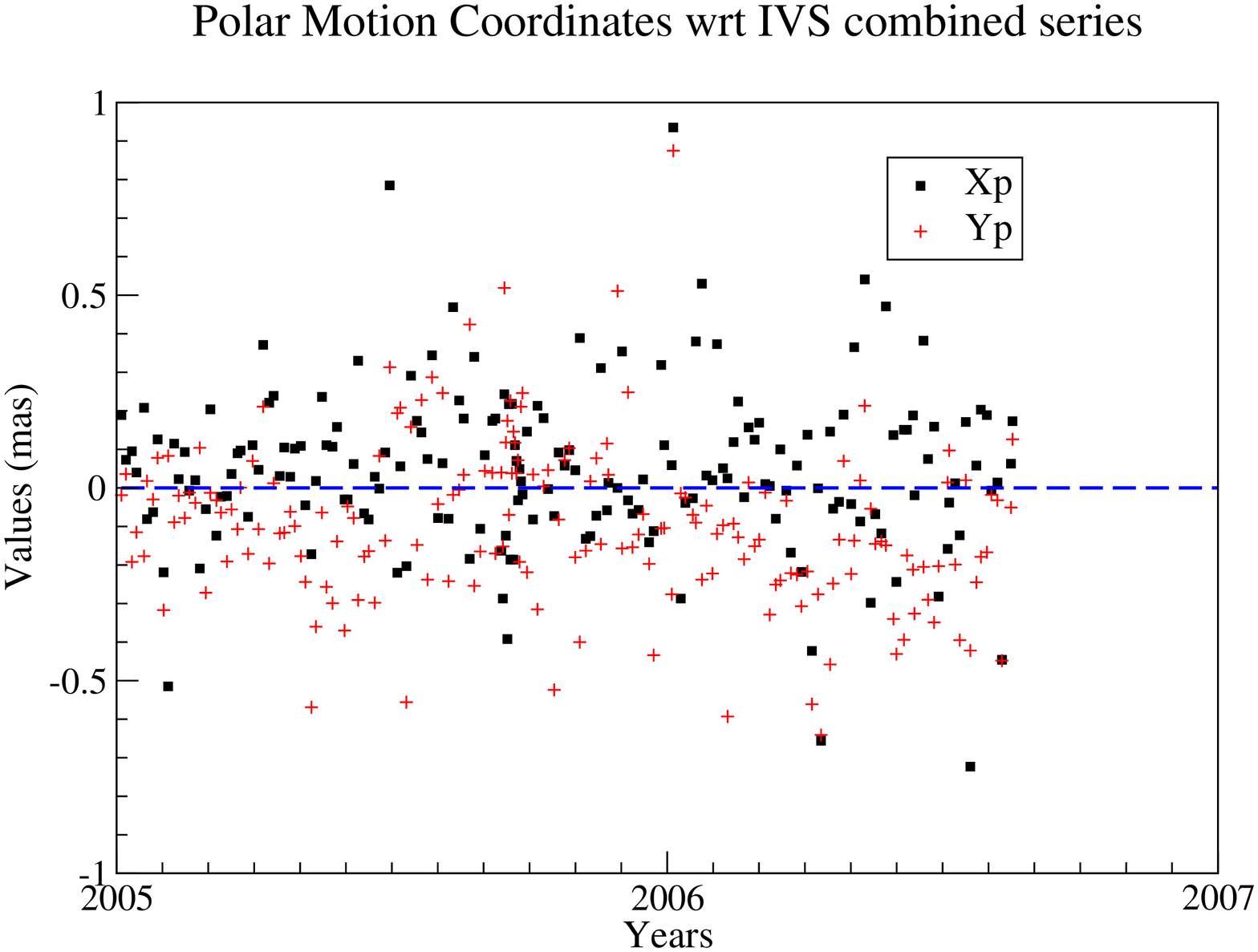}
  \includegraphics[scale=0.3,angle=-90]{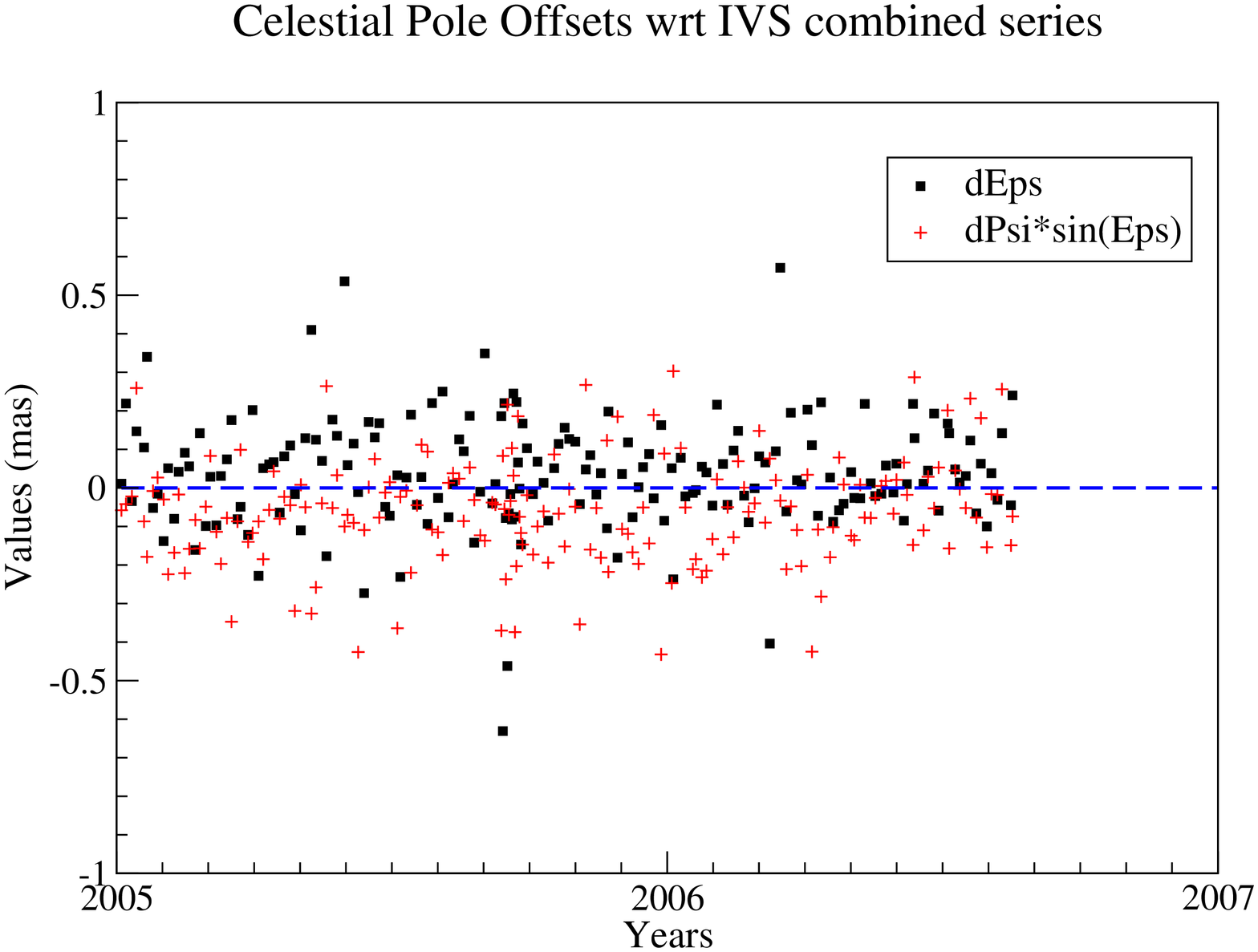}
  \includegraphics[scale=0.3,angle=-90]{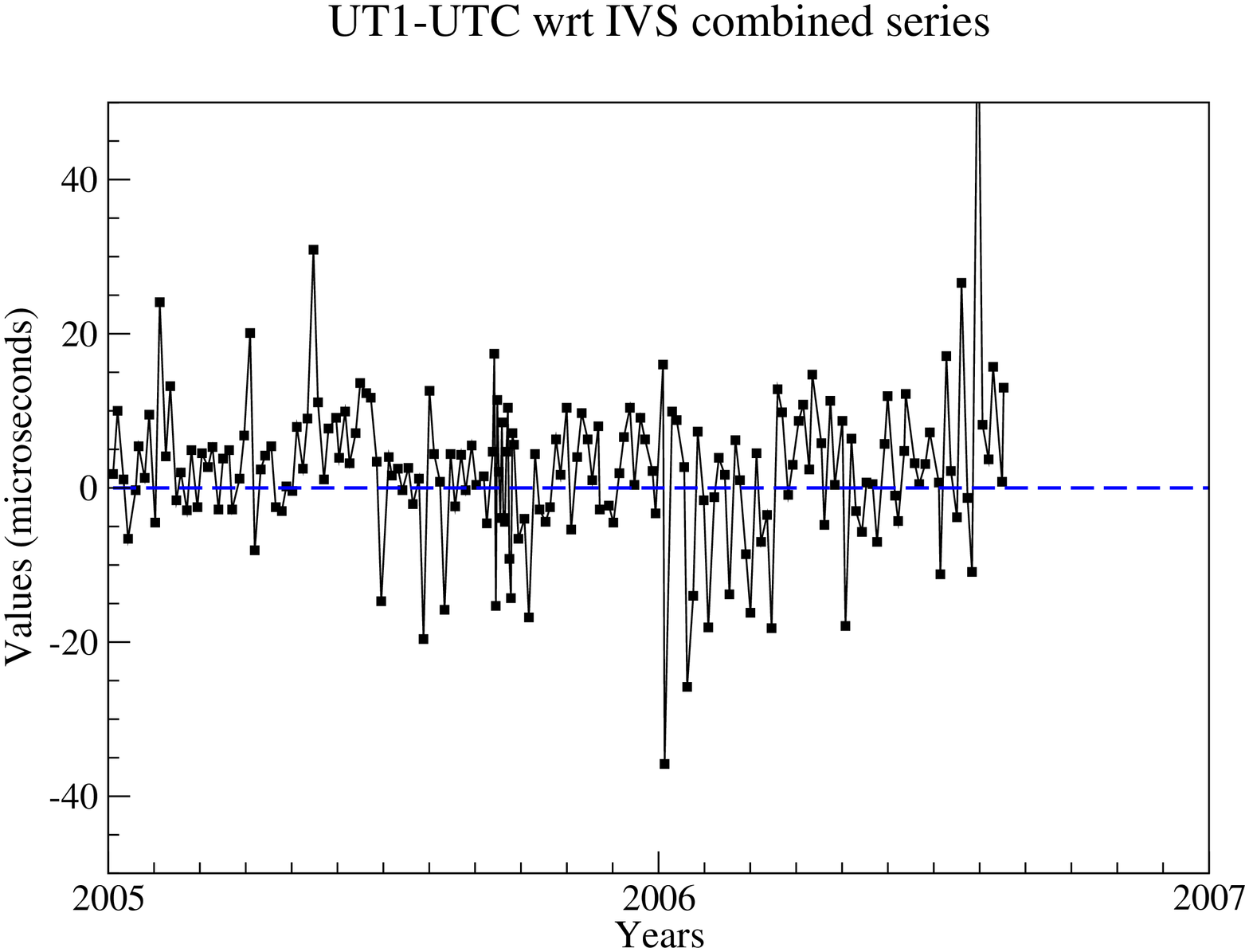}
\caption{VLBI Earth Orientation Parameters ($X_p$, $Y_p$, d$\epsilon$, d$\psi$ $\sin \epsilon$, $UT1-UTC$) estimated with GINS, compared to the IVS analysis coordinator combined series (ivs06q3e.eops) between 2005 and 2006.}
\label{fig:EOP}
\end{figure}

Figure~\ref{fig:EOP} shows the EOP series as derived with GINS, on the basis of the analysis described in the previous section. The results are plotted with respect to the IVS combined series (ivs06q3e.eops; see http://vlbi.geod.uni-bonn.de/IVS-AC/combi-eops/QUAT/HTML/start\_q.html). Table~\ref{tab:EOP_wrt_C04_IVS} summarizes the statistics for these series, plus those with respect to the IERS C04 series. Our current VLBI-only EOP results agree with the IVS combined series at the following levels (see RMS values in Table~\ref{tab:EOP_wrt_C04_IVS}):
\begin{itemize}
\item  0.20 mas for the polar motion coordinates, 
\item  0.15 mas for the celestial pole offsets, and 
\item  10 $\mu$s for the Earth's angle of rotation. 
\end{itemize}

These differences may partly arise from using different TRF in the two analyses: the IVS analysis coordinator used ITRF2000, whereas we fixed the TRF to VTRF2005.
Another point to highlight is that these comparisons used non-weighted RMS to evaluate the differences between the EOP series obtained with GINS and those published by the IVS analysis coordinator. Such values are generally larger that the more common ``weighted RMS''.

\begin{table*}
\centering
   \caption{Mean and RMS differences for each of the five EOP series ($X_p$, $Y_p$, $UT1-UTC$, d$\epsilon$, d$\psi$ $\sin \epsilon$) derived with GINS when compared to (1) the IERS C04 series, and (2) the IVS combined series.}
   \label{tab:EOP_wrt_C04_IVS}
   \begin{tabular}{llrrrrr}
      \hline \noalign{\smallskip}
                 &  EOP  &  $X_p$  &  $Y_p$  &  $UT1-UTC$  &  d$\epsilon$  &  d$\psi \sin \epsilon$ \\
                 &  wrt  & $\mu$as & $\mu$as &  $\mu$s     &  $\mu$as      &  $\mu$as               \\      
      \noalign{\smallskip} \hline \noalign{\smallskip}
      
      Mean       &  C04  & -201    &  343    &    3.4      &    49         &     7                 \\ 
      \noalign{\smallskip} 
                 &  IVS  &   47    &  -95    &    2.3      &    36         &   -62                 \\ 
      \noalign{\smallskip} \hline \noalign{\smallskip}
      RMS        &  C04  &  216    &  215    &    9.9      &   150         &   139                 \\ 
      \noalign{\smallskip} 
                 &  IVS  &  212    &  211    &   10.2      &   145         &   139                 \\  
      \noalign{\smallskip} \hline
    \end{tabular}
\end{table*}

%--------------------------------------------------------------------------------------
%--------------------------------------------------------------------------------------
\section{Conclusions and prospects}

GINS is a new VLBI analysis software in the IVS community. We showed that VLBI analyses undertaken with GINS lead to EOP results that agree at the level of 150--200 $\mu$as with respect to the IVS combined series results. This is comparable to the other IVS analysis centers, which demonstrates the capability of the GINS software for VLBI-only analysis. 

Another strength of GINS is the possibility of analysing observations of five space geodetic techniques (VLBI, GPS, LLR, SLR and DORIS) alltogether. Such a combination at the observation level is one of the goals of the IAG (International Association of Geodesy) project GGOS (Global Geodetic Observing System; \citealp{Rummel2005}).  

In the future, further developments and investigations are planned to refine such VLBI analysis with GINS:
\begin{itemize}
\item  To improve the relative weighting of the VLBI observations.
\item  To adjust tropospheric gradients, together with zenithal tropospheric delays.
\item  To validate the underlying models in GINS by carrying out detailed comparisons with those implemented in the JPL (Jet Propulsion Laboratory) VLBI software MODEST \citep{Sovers1996}.
\item  To submit to the IVS analysis coordinator the EOP results obtained with GINS for further evaluation.
\item  To adjust the radiosource coordinates to investigate the source position variability and ultimately to produce a celestial reference frame with GINS. 
\end{itemize}

%--------------------------------------------------------------------------------------
\subsection*{Acknowledgements}
G. Bourda is grateful to the CNES (Centre National d'Etudes Spatiales, France) for the post-doctoral position granted at Bordeaux Observatory. She wishes also to the Advisory Board of the Descartes-Nutations prize for supporting the journey to Vienna in order to present this work.

%--------------------------------------------------------------------------------------
%--------------------------------------------------------------------------------------

\end{document}